\documentstyle[12pt]{article}

\input{epsf}

\begin{document}
\begin{titlepage}
\begin{center}

\today    \hfill       MIT-CTP-2303
\hskip .5truein       \hskip .5truein hep-th/9408115\\

\vskip .2in

{{\Large \bf  Large  $N$ Phases of Chiral QCD${}_2$
\footnote{This work was supported in part by  funds provided by the
U.S. Department of Energy (D. O. E.)  under cooperative agreement
%DE-AC02-76ER03069,
DE-FC02-94ER40818
and in part by the divisions of Applied Mathematics of
the U.S. Department of Energy under contracts DE-FG02-88ER25065 and
DE-FG02-88ER25066.}}}

\vskip .15in

Michael Crescimanno\footnote{{\tt crescimanno@mitlns.mit.edu} \\
(After Sept.\ '94, {\tt
michael{\_}f.crescimanno@Berea.edu}, Physics Department,CPO 325
Berea College, KY 40404)} \\
Washington Taylor\footnote{\tt wati@mit.edu}\\

{\small {\em
Center for Theoretical Physics \\
Laboratory for Nuclear Science \\
and Department of Physics\\
Massachusetts Institute of Technology \\
Cambridge, Massachusetts 02139, USA}}
\end{center}
\vskip .15in

\begin{abstract}
A matrix model is constructed which describes a chiral version of
the large $N$ $U(N)$ gauge theory on a two-dimensional sphere of area
$A$.  This theory has three separate phases.  The
large area phase describes the associated chiral string
theory.  An exact expression for the free energy in the large area
phase is used to derive a remarkably simple formula for the number of
topologically inequivalent covering maps of a sphere with fixed branch
points and degree $n$.

\end{abstract}
\end{titlepage}
\newpage
\renewcommand{\thepage}{\arabic{page}}
\setcounter{page}{1}

%THIS IS PAGE 1 (INSERT TEXT OF REPORT HERE)
%\Large
\section {Introduction}
\setcounter{equation}{0}
\baselineskip 18.5pt

% Section 1
%

Although QCD is a relatively old subject, the  nonperturbative
calculation of physically significant
quantities in this theory is poorly understood.
Two-dimensional pure QCD, however, does admit a nonperturbative exact
solution [1-3].
%\cite{migdal,rusakov,witten}.
Recent work [4-8]
%\cite{gross,minahan,gt1,gt2,nrs}
has shown further that there is string theory description of pure
two-dimensional
gauge theories with various gauge groups.  It is hoped
that some modification of this string picture will  provide a
systematic nonperturbative description of QCD in any dimension.

A particularly interesting feature of two-dimensional QCD is that when
the theory is considered on a two-sphere, there is a phase transition
between the large and small area limits of the large $N$ gauge theory.
It was shown by Douglas and Kazakov \cite{dk} that the string
expansion is asymptotic to the full nonperturbative QCD$_2$ partition
function only when the dimensionless area is greater than $\pi^2$;
below this value, the theory simplifies dramatically and is
essentially a free theory described by a master field satisfying the
Wigner semicircle law.

The existence of this phase transition raises the question of whether
an analogous transition will occur in the four dimensional theory.
Such a phase transition would mean that a calculation in the strong
coupling regime would be unable to make useful predictions about
physics at weak coupling, so it is certainly important to understand
the nature of this type of phase transition.

Gross and Matytsin \cite{gm} have studied the DK phase transition from
the point of view of the weak coupling expansion.  They found that for
weak coupling (in two dimensions), there are nonperturbative instanton
contributions to
the partition function which vanish in the asymptotic expansion below
the critical point but which give rise to an infinite instanton
density above that point, driving the DK phase transition.  An
analogous calculation indicated that this mechanism
might not cause a phase transition in four dimensions.

The phase transition has also been studied from the
string point of view in
\cite{taylor}.  Analytic and numerical results indicate
that the string expansion has a radius of convergence which coincides
with the critical point.  In the string language, the transition is
driven by the entropy of branch point singularities in the string
maps.  (The formula for this entropy which was conjectured in
\cite{taylor} is proven in  this paper.)

Although these studies have given us some understanding of the reasons
for the phase transition in two-dimensional QCD, there are still many
unanswered questions.  Studying similar phase transitions in closely
related theories may give insights into the structure of these
transitions.

An interesting feature of the large $N$ QCD theory in
two dimensions is that the theory almost exactly factorizes into two
copies of a simpler, ``chiral'' theory \cite{gt1,gt2}. The chiral
theory encapsulates most of the important features in the geometry of
the underlying string theory, and seems worthy of study as a theory in
its own right.

The goal of this paper is to study
a chiral sector of the two-dimensional gauge theory on the sphere and to
determine its large $N$
phase structure.
We find that the chiral theory
contains 3 distinct phases, as opposed to the 2 phases of the
combined
theory.  Furthermore, we find that the same analytic
formulae describe both the large and small area limits; in analogy
with what was found in \cite{taylor}.
These formulae also reproduce
the asymptotic string expansion.
This suggests that the chiral theory, which is not yet
well understood from the point of view of the large $N$ gauge theory,
may have unexpected duality properties.

In Section 2 we briefly describe the QCD$_2$ string and review the
large $N$ theory and the Douglas-Kazakov phase transition.  We define the
chiral theory and give an Ansatz for the matrix model corresponding to
this theory.  Section 3 contains a saddle point evaluation of the
chiral partition function.  We find a class of single-cut saddle point
solutions
%(the so-called single-cut solutions)
which agree order by order with the string expansion and which are
defined for large and small areas, but not in an intermediate
region.  In Section 4, the single-cut solution is used to derive a
previously conjectured formula for the number of sphere to sphere maps
with $n$ branch points.  In Section 5 we find double-cut solutions for
the saddle point equations and show that these solutions interpolate
exactly between the regions of weak and strong coupling given by the
single-cut solutions.  Section 6 contains a discussion of our results
and related issues.

\section {Preliminaries}
\setcounter{equation}{0}
\baselineskip 18.5pt

The partition function of two-dimensional Yang-Mills theory
on a
compact manifold with genus $G$ and area $A$ is given by the exact
formula \cite{rusakov,witten}
\begin{eqnarray}
Z(G, A,N) & = &  \int[{\cal D} A^\mu]
{\rm e}^{- {1\over 4 g^2} \int_{\cal M}
 {\rm Tr} (F \wedge^{\Large \star} F)} \nonumber\\
  & = & \sum_{R} (\dim R)^{2 - 2G}
  {\rm e}^{-\frac{A}{2 N}C_2(R)}, \label{eq:partition}
\end{eqnarray}
where the sum is taken over all irreducible representations of the
gauge group, with $\dim R$ and $C_2(R)$ being the dimension and
quadratic Casimir of the representation $R$.  We absorb the coupling
constant $\lambda = g^2 N$ into the dimensionless area $A$ throughout
this paper.

We are  interested in studying the Yang-Mills theories with gauge
groups $U(N)$ and $SU(N)$.  The irreducible representations of these
gauge groups are associated with Young diagrams with row lengths
$n_i$, where for $U(N)$ the row lengths satisfy
\begin{equation}
\infty > n_1 \geq \cdots \geq n_N > -\infty
%\label{eq:}
\end{equation}
and for $SU(N)$
\begin{equation} \infty > n_1 \geq
\cdots \geq n_N = 0.
%\label{eq:}
\end{equation}
We denote the sum of
the row lengths by $n = \sum n_i $.  In terms of the row lengths, the
dimension of a representation $R$ is given by
\begin{equation}
 \dim R = \prod_{i > j} \left(1-\frac{n_i-n_j}{i-j}  \right).
%\label{eq:}
\end{equation}
For $U(N)$ the quadratic Casimir is given by
\begin{equation}
C_2 (R) = N n + \sum_{i} n_i (n_i-2i + 1),
%\label{eq:}
\end{equation}
and for $SU(N)$ we have
\begin{equation}
C_2 (R) = N n + \sum_{i} n_i (n_i-2i + 1)-\frac{n^2}{ N}.
%\label{eq:}
\end{equation}

In \cite{gt1,gt2} the asymptotic expansion in $1/N$ of the $SU(N)$
partition function (\ref{eq:partition}) was described in terms of a
string theory of covering maps.  This asymptotic expansion was
constructed by summing over a subset of representations called
``composite'' large $N$ representations; these are essentially the
representations whose quadratic Casimir has a leading term of order
$N$.  Composite representations are formed by taking the tensor
product of a representation $R$ corresponding to a Young diagram with
a finite number of boxes, and a representation $\bar{S}$ corresponding
to the complex conjugate of a representation $S$ associated with
another diagram with a finite number of boxes.  The resulting large
$N$ expansion of (\ref{eq:partition}) essentially factorizes into two
copies of a simpler, ``chiral'' string expansion.  The chiral
expansion arises from restricting the sum to only representations
corresponding to diagrams with a finite number of boxes.  The string
geometry of the chiral theory is simpler than that of the full theory
because in the chiral theory all maps from the string world sheet to
the target space have the same relative orientation, so the theory can
be viewed as a theory of orientation-preserving string maps.  The
chiral QCD${}_2$ string theory has recently been constructed by
Cordes, Moore, and Ramgoolam in terms of a topological string theory
\cite{cmr};  related work appears in \cite{horava}.

In the string expansion of (\ref{eq:partition}), a term of order
$N^{2-2g}$ in the partition function is interpreted as arising from a
(possibly disconnected)
string world sheet of genus $g$.  The dimension of a representation
associated with a Young diagram containing a finite number $n$ of
boxes has a leading term of order $N^n$.  Thus, when the genus $G$ is
greater than 1, the leading order nontrivial term in the chiral
partition function is of order $N^{2-2G}$ and arises from the
fundamental representation (with Young diagram
{\,\lower0.9pt\vbox{\hrule
\hbox{\vrule height 0.2 cm
\hskip
0.2 cm \vrule height 0.2 cm}\hrule}\,}).  In the full $SU(N)$ theory,
both the fundamental representation and its conjugate contribute terms
of order $N^{2-2G}$.  Because in these cases only a finite number of
terms contribute to the leading order term in the partition function,
the large $N$ theory with $G > 1$ clearly cannot have a phase
transition at leading order.  When $G = 1$, the contribution of the
dimension factor to the partition function vanishes and so all
representations contribute to the $N^0$ term.  However, in this case
the contribution of all the terms can easily be summed and found to
give a theory with no phase transitions at leading order.  The leading
term in the free energy of this theory corresponds in the string
picture to a summation over unbranched covers of a torus by a torus.

For the spherical case $G = 0$, however, the situation is more
complicated.  In the partition function, terms corresponding to Young
diagrams with $n$ boxes give rise to terms of order $N^{2n}$.  The
free energy of the string expansion, given by the logarithm of the
partition function, has a leading term of order $N^2$.  However, an
infinite number of string diagrams contribute to the coefficient of this
leading term, and so a phase transition in the leading order
term cannot be ruled out.

The possibility of a phase transition in the genus 0 free energy was
investigated in
\cite{rusakov2,dk,mp}.  These papers considered the $U(N)$ theory,
using the continuum variables
\begin{equation}
x = \frac{i}{ N}  \; \; \;\; \; \;
n ( x) = \frac{n_i}{ N}
%\label{eq:}
\end{equation}
to describe Young diagrams in the large $N$ limit.  Changing variables
to
\begin{equation}
h (x) = -n (x) + x-1/2,
%\label{eq:}
\end{equation}
the partition function can be rewritten as a functional integral
\begin{equation}
Z (A) = \int {\cal D} h (x) \;
{\rm e}^{-N^2 S[h (x)]},
%\label{eq:}
\end{equation}
where
\begin{equation}
S[h  (x)] =
 -\int_{0}^{1} {\rm d} x \int_{0}^{1} {\rm d}  y\;
\ln[h (x)-h (y)]  +\frac{A}{2}  \int_{0}^{1} {\rm d} x \;h(x)^2
-\frac{A}{24}.
\label{eq:action}
\end{equation}
This action is familiar from Hermitian
matrix model theories, where the log term
arises from the Van der Monde determinant factor associated with
integration over the angular degrees of freedom in the matrix.
The saddle point of this action is described by the equation
\begin{equation}
\frac{A}{2}  h = P \int {\rm d} s \; \frac{ u (s)}{h-s},
\label{eq:integral1}
\end{equation}
where $u (h)= {\rm d} x/{\rm d} h$. Since $x$ ranges
between $0$ and $1$, $u$ is normalized so that $\int u
= 1$.  Because the variables $n_i$ are nonincreasing, $u$ must satisfy
the constraint $u (h)\leq 1$.

The integral equation (\ref{eq:integral1}) has a solution given by the
Wigner semicircle law.  However, for values of $ A >
\pi^2$, the Wigner semicircle solution violates the constraint $u \leq
1$.  In order to deal with this problem, Douglas and Kazakov (DK)
adopted the Ansatz that the function $u (h)$ should be identically
equal to 1 in a region $[-b,b]$, and should be between 0 and 1 in the
regions $(-a,-b)$ and $ (b,a)$ for some $0<b<a$.  This Ansatz was
physically motivated by the idea that above the critical area, the
density function $u$ would become saturated to its maximum value of 1
in a region symmetric around the origin, due to the symmetry of the
quadratic potential.  Using this Ansatz, DK solved the integral
equation (\ref{eq:integral1}) in the nonconstant regions of $u$, and
found an expression for $u (h)$ and the leading term in the derivative
of the free energy in terms of complete elliptic functions.  They
found a third order phase transition separating the large and small
$A$ phases at $A = \pi^2$, and verified that the free energy in the
large area phase agrees with the string expansion.

In subsequent work \cite{mp}, it was pointed out by Minahan and
Polychronakos (MP) that the symmetric form of the DK Ansatz
effectively restricts the saddle point to the $Q = 0$ $U(1)$ charge
sector of the $U(N)$ theory.  They considered a more general
asymmetric Ansatz, in which $u$ has support in the region $(d,a)$, and
is identically 1 in the region $[c,b]$ where $d<c<b<a$.  By solving
the integral equation (\ref{eq:integral1}) using this Ansatz, they
found a more general set of solutions for the $U(N)$ theory, where for
a fixed value of $A$ there is a range of solutions corresponding to
different values of the $U(1)$ charge $Q$.  MP did not ascertain
whether the value of the free energy was larger or smaller on these
additional solutions, so that it is not clear whether these other
possible saddle points affect the free energy of the theory.  However,
it was shown in
\cite{taylor} that the free energy of the complete $SU(N)$ theory is
equal to the free energy of the $Q = 0$ $U(N)$ theory in the large
area phase where the string expansion is valid.  Thus, the results of
DK can be taken to be a complete description of the phase structure of
the large $N$ $SU(N)$ theory on the sphere.  This result seems to
indicate that in fact, for a fixed $A$,
the DK saddle points represent the extremum of
the action with respect to $Q$.
%We discuss this contention
%for our problem more fully in Section 6.

In this paper we analyze the chiral version of the large $N$ theory
using the matrix model approach.  The precise theory we will
investigate is the chiral version of the $Q = 0$ $U(N)$ theory, which
corresponds to a string theory of orientation-preserving maps with
branch points but no tube or handle singularities.  To define a matrix
model for the chiral theory, we choose an Ansatz for which $u$ selects
representations with row lengths $ n_i = 0$ for all $i$ greater than
some fixed value $k$.  In the continuum variables, this condition
corresponds to
\begin{equation}
u (h) = 1, \; \; \; \; \; {\rm for} \;c \leq h < 1/2,
\label{eq:condition}
\end{equation}
where $u (h)$ has support on the interval $(d, 1/2)$ for some $d \leq
c<1/2$.  The remainder of this paper is devoted to using
(\ref{eq:condition}) to calculate the saddle point and determine the
phase structure of the chiral theory.

\section {Single-cut Solution}
\setcounter{equation}{0}
\baselineskip 18.5pt

The simplest Ansatz for a function $u (h)$ satisfying
(\ref{eq:condition}) is that $u <1$ in the region $(d,c)$.  This will
result in a solution that is the analog of the Wigner semicircle
saddle point solution in the low area phase of the coupled theory
\cite{dk}.  Surprisingly we will find that it describes both the small
and large area phases of the chiral theory.  This Ansatz gives rise to
the integral equation
\begin{equation}
\frac{A h}{2}  + \ln \frac{h-1/2}{h- c} = P \int_{d}^{c}  {\rm d} s \;
\frac{ u (s)}{h-s}.
\label{eq:integralequation1}
\end{equation}
The general approach to solving an integral equation of this type
\cite{pipkin} is to define a function
\begin{equation}
f (h) = \int_{d}^{c} {\rm d} s \;
\frac{u (s)}{h-s}.
\label{eq:fdefn1}
\end{equation}
Note that $f$ is analytic
on the complex $h$ plane, with a cut along the interval $[d,c]$.
Defining $f_+ (h)$ and $f_- (h)$
to be the limiting values of $f$  as one approaches the cut from above
and below respectively, we have
\begin{equation}
{\cal F} (h) \equiv \frac{A h}{2}  + \ln \frac{h-1/2}{h- c} = 1/2 (f_+ +
f_-)
%\label{eq:}
\end{equation}
and
\begin{equation}
u (h) = \frac{-1}{2 \pi i}  (f_+ -f_-).
%\label{eq:}
\end{equation}
Let $g (h)$ be a function which is valued on the cut plane,
and let $g_+(h)$ and $g_-(h)$ be the limiting values of $g(h)$ as
one approaches the cut from above and from below respectively.
We choose a $g$ which satisfies $g_+ + g_-= 0$.  It follows that
\begin{equation}
-\pi i\frac{u (h)}{g_+ (h)}  = \frac{1}{ 2}
\left[(f/g)_+ + (f/g)_- \right]
%\label{eq:}
\end{equation}
and
\begin{equation}
-\frac{(f/g)_+ -(f/g)_-}{ 2 \pi i}
  = -\frac{f_+ + f_-}{2 \pi i g_+}  = \frac{-1}{\pi i} \frac{{\cal F}}{g_+}
{}.
%\label{eq:}
\end{equation}
But this indicates that
%(because these functions are in the same
%relation as $F$ and $u$ above)
\begin{equation}
\frac{f (h)}{g (h)} =  \int_{ d}^{ c}
{\rm d} s \;\frac{-1}{\pi i}  \frac{ {\cal F} (s)}{ (h-s)g_+ (s)}.
%\label{eq:}
\end{equation}
For our problem, a simple choice
for a function $g$ satisfying $g_+ + g_- = 0$ is
\begin{equation}
g (h) = \sqrt{(h-c)(h-d)},
\label{eq:gdefn1}
\end{equation}
and thus
\begin{equation}
f (h) = \frac{g (h)}{ 2 \pi i} \oint {\rm d} s \;
\frac{{\cal F} (s)}{(h-s)g (s)}
 = \frac{1}{2 \pi i}  \sqrt{(h-c)(h-d)}
\oint {\rm d} s \; \frac{\frac{As}{2}+\ln \frac{s-1/2}{ s-c}  }{
(h-s)\sqrt{(s-c)(s-d)}},
%\label{eq:}
\end{equation}
where the integral is taken around the cut $(c,d)$.
Deforming the contour, we get contributions from the poles at $h$ and
$\infty$ and the integral around the log cuts,
\begin{equation}
f (h) = \frac{Ah}{2}  +\ln ( \frac{h-1/2}{ h-c} )
-\frac{A}{2} \sqrt{(h-c)(h-d)}
+ \int_{c}^{1/2} {\rm d} s
\frac{\sqrt{(h-c)(h-d)}}{ (h-s)\sqrt{(s-c)(s-d)}}.
\label{eq:f}
\end{equation}
We can now calculate $u (h)$ in the interval $(d,c)$
\begin{eqnarray}
u (h)  &=  &\frac{1}{\pi}  \left[
\frac{A}{2}\sqrt{(c-h)(h-d)}
- \int_{c}^{1/2} {\rm d} s
\frac{\sqrt{(c-h)(h-d)}}{ (h-s)\sqrt{(s-c)(s-d)}}
  \right] \nonumber\\
& = &\frac{1}{ \pi}\left[
 \frac{A \sqrt{(c-h)(h-d)}}{2} +\tan^{-1} \left( -
\frac{\sqrt{(c-h)(h-d)(1-2c)(1-2d)}}{h-(h+ 1/2) (c + d)+ 2cd}  \right)
  \right]
\end{eqnarray}
where we take values of $\tan^{-1}$ to be between 0 and $\pi$.

The variables $c$ and $d$ are of course not arbitrary.
Since $u$ is normalized to $\int u (h)\; {\rm d} h = 1$, we know
that for large $h$, $f (h)$ satisfies
\begin{equation}
f (h) = \frac{1-1/2 + c}{h}  + (\cdots) h^{-2}   +
(-2 F' (A) + 1/24 + c^3/3)h^{-3}+
{\cal O} (h^{-4}).
\label{eq:expansion}
\end{equation}
where we note that $F(A)$ is the free energy, $F(A) = \ln(Z(A))/N^2$.
Expanding
(\ref{eq:f}) in powers of $h$ and performing the necessary
integrations, we find that $c$ and $d$
satisfy the conditions
\begin{equation}
-\frac{A}{4}(c + d)  = \ln\left( \frac{\sqrt{(1-2c)(1-2d)} +
1-d-c}{c-d} \right),
\label{eq:constraint1a}
\end{equation}
\begin{equation}
1 = A \frac{(c-d)^2}{ 16}  + \sqrt{(1/2 -c)(1/2 -d)}.
\label{eq:constraint1b}
\end{equation}

\begin{figure}
\epsfbox{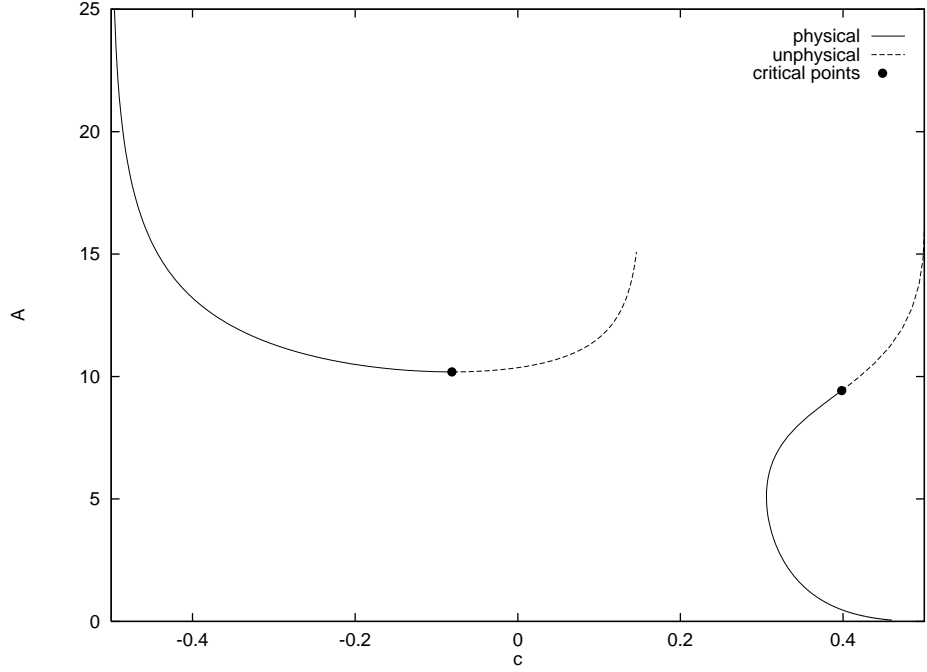}
\caption[x]{\footnotesize single-cut solutions}
\label{f:ca}
\end{figure}

In Figure~\ref{f:ca} we have graphed the solutions to
(\ref{eq:constraint1a}) and (\ref{eq:constraint1b}) on the $c$-$A$
plane.
These solutions lie on two disconnected curves.
We observe that in the large area regime, as $A
\rightarrow \infty$, we have $c,d \rightarrow -1/2$.  As $A
\rightarrow 0$, on the other hand, we have $c\rightarrow 1/2$, $d
\rightarrow -\infty$.
In section 4 we show that the large $A$ solution to these equations
reproduces the string solution.

%As we discuss in the next section, we find that
%the large area phase exactly reproduces the result of the string
%expansion.
%; the small area phase, however, seems more mysterious. It
%does not seem to correspond to the small area phase of the string
%expansion as strong similarity with the results of \cite{taylor} might
%suggest.   We will comment further on this issue in Section 6.

For a solution of the constraint equations to
give a physically acceptable saddle point we must further verify that
$u (h)\leq 1$.  Unphysical solutions are denoted in Figure~\ref{f:ca}
by  dashed lines.
Checking the solutions in the large  $A$ region, we find
that for sufficiently large $c$, the function $u$  is larger than 1
for $h \sim c-$.  The $A$ value for which
this first occurs is the point where
${\rm d} u/{\rm d} h = 0$ at $h = c$.
Furthermore, it is straightforward to verify that this is
precisely the point where ${\rm
d} A/{\rm d} c = {\rm d} d/{\rm d} c = 0$.
This implies that $c$ and $d$ satisfy the relation
\begin{equation}
4 \sqrt{\frac{{1/2-c}}{{1/2 -d}} }
+ 3c + d-2 = 0.
\label{eq:ucp}
\end{equation}
%where
%\begin{equation}
%x = \sqrt{1/2 -c}  \; \; \; \;
%y = \sqrt{1/2 -d}.
%\label{eq:}
%\end{equation}
Combining this with (\ref{eq:constraint1a}) and (\ref{eq:constraint1b})
we find that the area at this point is approximately
\begin{equation}
A_+  \approx 10.189
%\label{eq:}
\end{equation}

\begin{figure}
\epsfbox{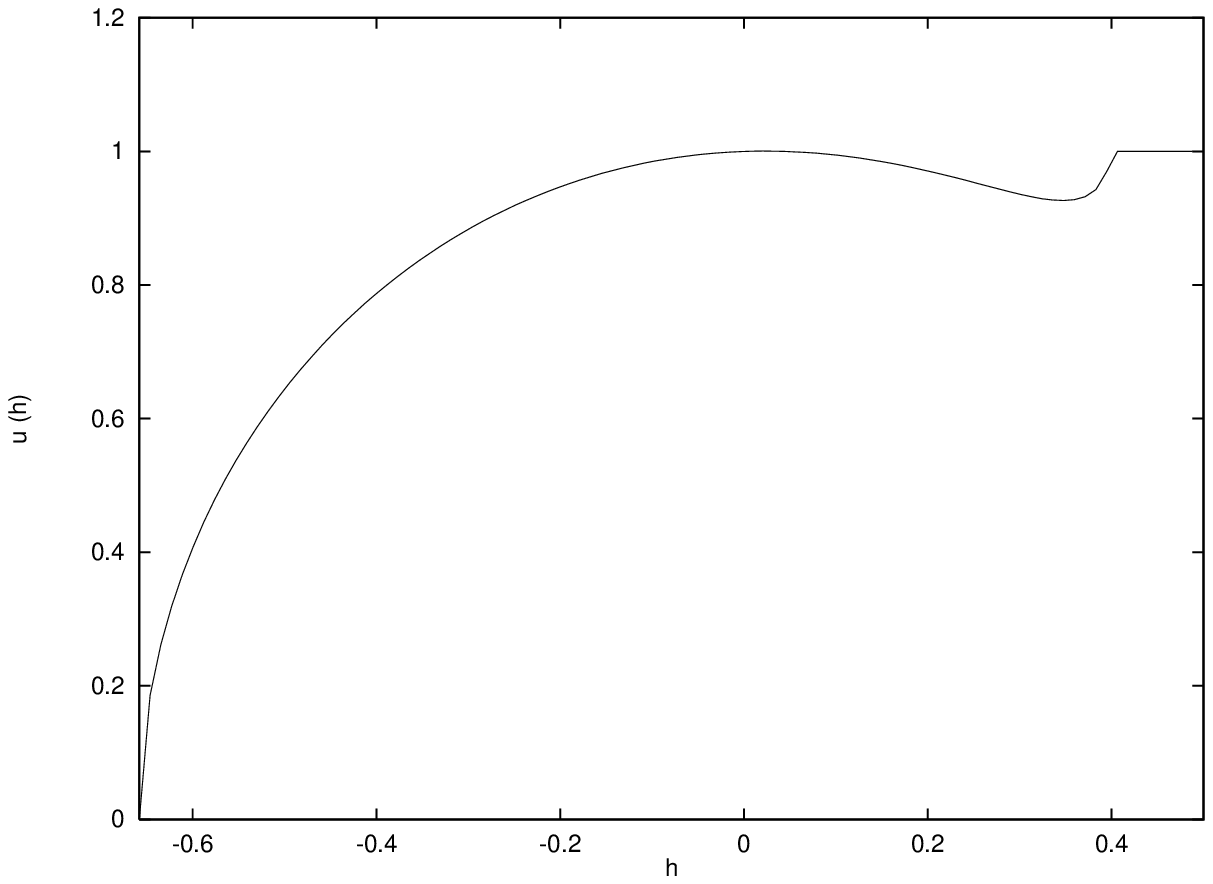}
\begin{center}
$A = A_+$ \\
\end{center}
\epsfbox{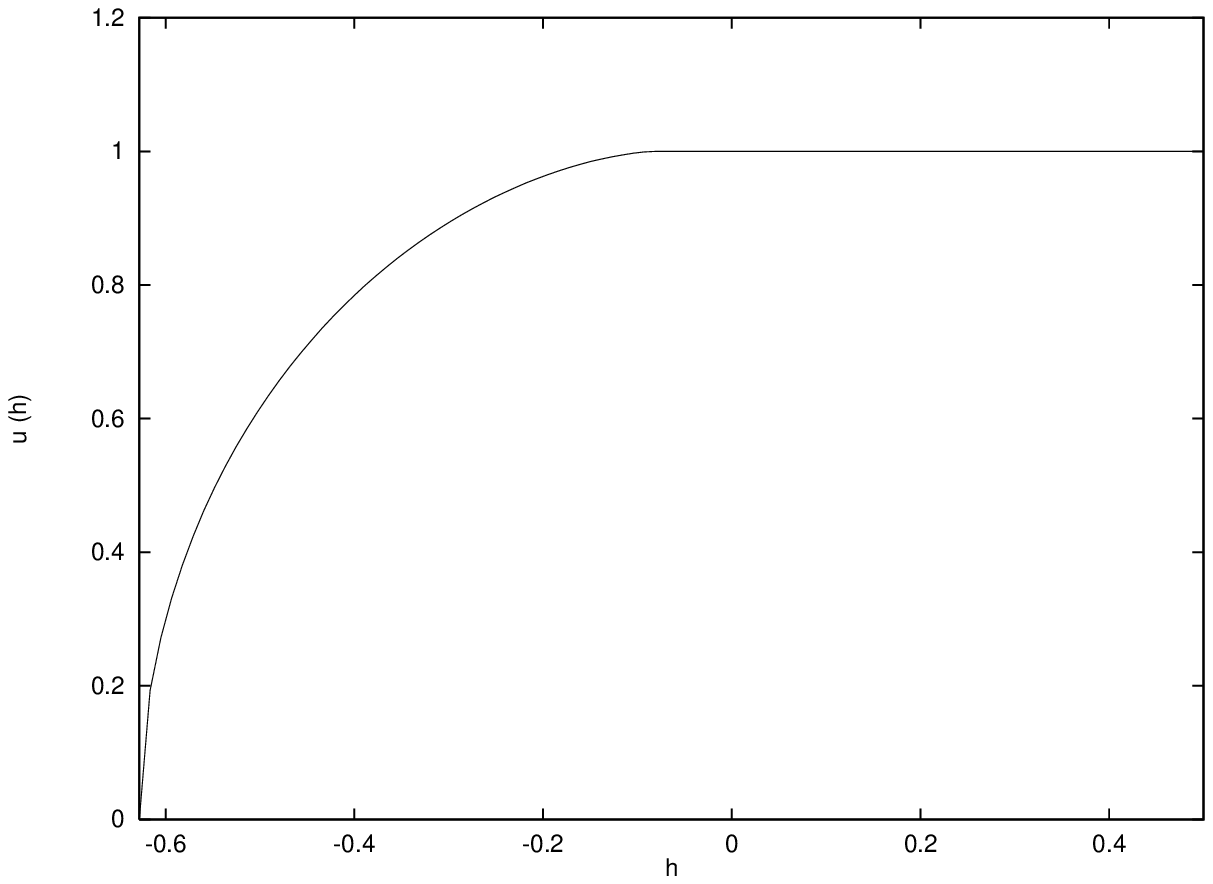}
\begin{center}
$A = A_-$ \\
\end{center}
\caption[x]{\footnotesize Function $u(h)$ at critical points}
\label{f:u}
\end{figure}

Similarly, we find that for sufficiently large $A$, the solutions on
the small $A$ curve  in Figure~\ref{f:ca} are unphysical because the
function $u$ develops a ``bump'', where $u (h) > 1$.  The
boundary of the physical region occurs at the point where the
equations
\begin{equation}
u (e) = 1 \; \; \; \;u' (e) = 0
%\label{eq:}
\end{equation}
have a simultaneous solution for some $d < e < c$.  These equations
simplify to the simultaneous equations
\begin{eqnarray}
\cos \left(\frac{A \sqrt{(c-e)(e-d)}}{4}\right) & = &
\sqrt{\frac{(e-d)( 1/2-c)}{ (c-d)( 1/2-e)}} \label{eq:lcp1}\\
\frac{\sqrt{(1/2 -c)(1/2 -d)}}{ 1/2 -e}
 & = &  \frac{A}{4}  (2e-c-d)
\label{eq:lcp2}
\end{eqnarray}
Again combining with (\ref{eq:constraint1a}) and (\ref{eq:constraint1b}),
the solution to these equations corresponds to an area of
approximately
\begin{equation}
A_- \approx 9.426
%\label{eq:}
\end{equation}
In Figure~\ref{f:u}, we have graphed  the function $u (h)$ at the two
values of area $A_+$ and $A_-$.
These represent the boundaries of the physical region of the
single-cut solution.

Given the solutions we have found to the integral equation
(\ref{eq:integralequation1}) which satisfy the constraint $u (h)\leq
1$, we can use (\ref{eq:f}) and
(\ref{eq:expansion}) to calculate $F' (A)$, the
derivative of the free energy with respect to the area,
for the solutions of the
single-cut Ansatz.  We have
\begin{eqnarray}
-2 F' (A) & = & \left[-1/12
+\frac{(c-d)^2 (5c^2 + 6cd + 5d^2)A}{256}  \right.  \\
& &\hspace{0.3in}\left. + \int_{c}^{1/2} {\rm d} s
\frac{s^3-s^2 (c + d)/2-s (c-d)^2/8-(c + d) (c-d)^2/16}{\sqrt{(s-c)(s-d)}}
\right].\nonumber
\end{eqnarray}
Performing the integration and using
(\ref{eq:constraint1b}), we have
\begin{equation}
F' (A) =
\frac{A (c-d)^2}{384} (1 + c + d    -
\frac{9 c^2  + 14 cd + 9 d^2}{4})
- (\frac{c + d}{24}  +
\frac{ 3 c^2  +2 cd +  3 d^2}{48}).
\label{eq:free}
\end{equation}

We have thus found that for any area outside the region $(A_-,A_+)$
there is a unique solution to the integral equation
(\ref{eq:integralequation1}) which satisfies $u(h) < 1$ for $d < h <
c$.  We have calculated the derivative of the free energy for these
solutions.  However, for $A_-< A < A_+$ we have found no solution
corresponding to a function $u$ representing a master field for the
large $N$ Young diagram.  In Section 5, we will show that allowing $u$
to be identically 1 in another interval provides a solution in the
intermediate area regime.

\section {Enumeration of Sphere Maps}
\setcounter{equation}{0}
\baselineskip 18.5pt

Now that we have an analytic expression for the derivative of the free
energy of the chiral theory for $A > A_+$ and for $A < A_-$,  we
expect that in the large area regime this formula should agree with
the  free energy of the chiral QCD${}_2$ string.  The chiral string
theory has a free energy which is expressed as a double expansion in
${\rm e}^{-A/2}$ and $A$.
Specifically, the free energy of the chiral theory  can be written
\cite{taylor}
\begin{equation}
\sum_{n = 1}^{\infty}   \phi_n (A)  x^n,
%\label{eq:}
\end{equation}
where
\begin{equation}
x ={\rm e}^{-\frac{A}{2} },
%\label{eq:}
\end{equation}
and the functions $\phi_n (A)$ are polynomials of degree $2n-2$ in
$A$.  To check the correspondence with (\ref{eq:free}) we must expand
our solutions for $c,d,$ and $F'$ as double expansions around $A =
\infty$.  Writing
\begin{equation}
c = -1/2 + \epsilon \; \; \; \; d = -1/2 -\delta,
%\label{eq:}
\end{equation}
the conditions (\ref{eq:constraint1a}, \ref{eq:constraint1b}) on $c,d$
become
\begin{equation}
{\rm e}^{-A/4} {\rm e}^{(\epsilon -\delta)A/4} =
\frac{\epsilon + \delta}{2 (1 + \sqrt{(1-\epsilon)(1 + \delta)})
+ \delta -\epsilon}
%\label{eq:}
\end{equation}
and
\begin{equation}
1-\sqrt{(1-\epsilon)(1 + \delta)}= \frac{A (\epsilon + \delta)^2}{16}.
%\label{eq:}
\end{equation}
We can rewrite these two equations as
\begin{equation}
\epsilon + \delta
=  \sqrt{x} \left(4-\frac{A (\epsilon + \delta)^2}{8}  + \delta -\epsilon
\right) {\rm e}^{(\epsilon -\delta)A/4}.
\label{eq:recursive1}
\end{equation}
and
\begin{equation}
\delta -\epsilon = \delta \epsilon -
\frac{A (\delta + \epsilon)^2}{8}
+\frac{A^2 (\delta + \epsilon)^4}{256}.
\label{eq:recursive2}
\end{equation}
These equations can be solved order by order in $\sqrt{x}$,  giving
$c$ and $d$ as power series' in $x^{1/2}$ with the coefficients of
$x^{n/2}$ being polynomials in $A$ of order $n-1$.  Computing up to $n
= 2$, for example, we have
\begin{eqnarray}
c & = &-1/2 + 2x^{1/2}+ (A-2)x
+ (A^2 -4A + 2)x^{3/2}\nonumber\\
& &\hspace{1in}+(A^3-6A^2 + 8A-2) x^2 +
 {\cal O}(x^{3/2})  \nonumber \\
d & = & -1/2 - 2x^{1/2} + (A-2)x
- (A^2 -4A + 2)x^{3/2}\\
& &\hspace{1in}+(A^3-6A^2 + 8A-2) x^2 + {\cal O}(x^{3/2}). \nonumber
\end{eqnarray}
Plugging  these expansions into
(\ref{eq:free}), we have an expansion for $F' (A)$ in $x$ and $A$.
The lowest order terms in this expansion give
\begin{equation}
F' (A) =- \frac{1}{2}  x -(\frac{A^2}{4} -\frac{3A}{2}  + \frac{3}{2}
)x^2 + \cdots
%\label{eq:}
\end{equation}
We have calculated this expansion out to order $x^{12} $, and find exact
agreement with the coefficients in the polynomials $\phi_n$ of the
string expansion.  The polynomials $\phi_n (A)$ for $n \leq 7$ are
given explicitly in \cite{taylor}.

Thus, we have verified that in the large area regime, the matrix model
we have constructed correctly reproduces the string theory of chiral
QCD${}_2$.  We now have an exact formula for the free energy of this
theory, which we can use to calculate to arbitrary order the terms in
the expansion.  A particularly interesting calculation is the
determination of the leading coefficients in the polynomials $\phi_n
(A)$.  For a fixed value of $n$, the polynomial $\phi_n$ is of degree
$2n-2$, and the coefficient $\gamma_n$ of $A^{2n-2}/(2n-2)!$ is equal
to the number of topologically distinct holomorphic maps from $S^2$ to
$S^2$ with $2n-2$ elementary branch point singularities whose images
are fixed.  It was conjectured by Gross and Taylor \cite{taylor} that
these coefficients are given by
\begin{equation}
\gamma_n = \frac{n^{n-3}(2n-2)!}{n!}
\label{eq:conjecture}
\end{equation}
We can now use the exact expression for the free energy
(\ref{eq:free}) to prove that these are precisely the leading
coefficients in the matrix model free energy expansion, and thus by
implication that the coefficients $\gamma_n$ of the QCD${}_2$ string
are indeed given by
(\ref{eq:conjecture}).

To calculate the desired coefficients in the free energy
expansion, it will be useful to consider a simplification of the
theory described above in which subleading terms in the polynomial
coefficients of powers of $x$ are dropped.  We denote the free energy
of the simplified theory by $\tilde{F}$.
This free
energy has a derivative which is of the form
(continuing to drop subleading terms)
\begin{equation}
\tilde{F}' (A) = \sum_{n} \zeta_n  A^{2n-2} x^n.
%\label{eq:}
\end{equation}
In the string theory language, this simplification corresponds to
removing $\Omega$-points from the theory.
{}From the form of equations
(\ref{eq:recursive1}) and (\ref{eq:recursive2}), it is straightforward
to show that the expansions of $\epsilon$ and $\delta$ are of the form
\begin{eqnarray}
\epsilon & = &  \sum_{n=1}^{\infty}
x^{n/2}(e_nA^{n-1}+ {\cal O}(A^{n-2}))\\
\delta & = &  \sum_{n=1}^{\infty}
x^{n/2}(d_nA^{n-1}+ {\cal O}(A^{n-2})).
\end{eqnarray}
Rewriting (\ref{eq:free}) in terms of $\epsilon$ and $\delta$, and
dropping terms which are of lower order in $A$ for each power of $x$,
we have
% insert ~after this
\begin{equation}
\tilde{F}' (A) =
 \frac{2 (\tilde{\epsilon} -\tilde{\delta})^2
- (\tilde{\epsilon} + \tilde{\delta})^2}{16}
= \frac{1}{8} (\tilde{\epsilon} -\tilde{\delta})^2 -\frac{1}{2A}
(\tilde{\epsilon} -\tilde{\delta}).
\label{eq:free2}
\end{equation}
where
\begin{eqnarray}
\tilde{\epsilon} & = &  \sum_{n=1}^{\infty}
e_n A^{n-1} x^{n/2}\\
\tilde{\delta} & = &  \sum_{n=1}^{\infty}
d_n A^{n-1} x^{n/2}.
\end{eqnarray}
Dropping  lower order terms in equations (\ref{eq:recursive1}) and
(\ref{eq:recursive2}), we have
\begin{equation}
\tilde{\epsilon} + \tilde{\delta} =
4x^{1/2} {\rm e}^{(\tilde{\epsilon} -\tilde{\delta})A/4}
%\label{eq:}
\end{equation}
and
\begin{equation}
\tilde{\epsilon} -\tilde{\delta} =
\frac{A (\tilde{\delta} + \tilde{\epsilon})^2}{8}.
%\label{eq:}
\end{equation}
Combining these equations,
\begin{equation}
\xi =\frac{A (\tilde{\epsilon} -\tilde{\delta})}{2} = A^2 x {\rm e}^{\xi}.
\label{eq:xi}
\end{equation}

We will now find it useful to recall some properties of the generating
function
\begin{equation}
f (z) = \sum_{n = 1}^{\infty} \frac{n^{n-2}}{(n-1)!}z^n.
%\label{eq:}
\end{equation}
A famous theorem due to Cayley \cite{cayley} asserts that $c_n
=n^{n-2}$ is the number of distinct connected graphs with $n$ labeled
vertices and $n -1$ edges.  If we remove a fixed vertex from every
graph of this type, and sum over all possible graphs by summing over
the number $k$ of edges connected to that vertex and over all possible
subgraphs connected to each edge, we can derive the recursive equation
\begin{equation}
c_n =\sum_{k}  \sum_{m_1 + \cdots +m_k = n-1}
\frac{ (n-1)!}{k!m_1!\cdots m_k!} \prod_{i=1}^{k} m_i c_{m_i}.
%\label{eq:}
\end{equation}
This equation implies that $f (z)$ satisfies
\begin{equation}
f (z) = z {\rm e}^{f (z)}
\label{eq:identity1}
\end{equation}
If we were to label edges instead of vertices, the number of distinct
graphs would be $c_n/n$ (where graphs with symmetry are counted with a
weight of the inverse of the order of symmetry).  By removing a fixed
edge from each graph, and counting all graphs connected to that edge
as above, we arrive at the equation
\begin{equation}
c_n/n = 1/2 \sum_{i = 1}^{n-1}
\frac{  (n-2)!}{(i -1)!(n-i-1)!} c_i c_{n-i}
%\label{eq:}
\end{equation}
This gives rise to the equation for the generating function,
\begin{equation}
f (z)^2 = 2\sum_{n =1}^{\infty}  \frac{n^{n- 3}}{(n-2)!}  z^n.
%\label{eq:}
\end{equation}

{}From (\ref{eq:identity1}) and (\ref{eq:xi}), we see that
\begin{equation}
\xi = f (A^2x).
%\label{eq:}
\end{equation}
We can now  rewrite the derivative of the free energy in the form
\begin{equation}
\tilde{F}' (A) =
 \frac{f (A^2x)^2}{2A^2}  -\frac{f (A^2x)}{A^2}
= -\sum_{n = 1}^{\infty} \frac{n^{n-3}}{ (n-1)!}  A^{2n-2} x^{n}.
%\label{eq:}
\end{equation}
This is precisely the desired result, and proves that the coefficients
in (\ref{eq:conjecture}) describe the free energy for the large area
phase of the matrix model described in the previous section.  The
equivalence of this matrix model with the chiral string expansion of
QCD${}_2$ indicates that the number of topologically distinct degree
$n$ covers of the sphere by a sphere with $2n-2$ fixed branch points
is exactly $n^{n-3}(2n-2)!/n!$.  This result was used (but not proven)
in \cite{taylor} to argue that in the strong coupling string expansion
the entropy of branch point singularities drives the phase transition
on the sphere, both in the chiral and coupled theories.

\section {The Double-Cut Solution}
\setcounter{equation}{0}
\baselineskip 18.5pt

% Double-Cut
%

In Section 3, we derived a single-cut
solution to the equation of motion for both small
($A<A_{-}$) and large areas ($A>A_{+}$). For intermediate areas
$A_{-}<A<A_{+}$ there are no physically viable single-cut
solutions.
This is due to the fact that the functions $u(h)$  describing the
single-cut solution for the areas in this range are ``unphysical'';
they do not satisfy the constraint $u(h)<1$
and thus cannot correspond to Young diagrams.

This situation is analogous to that which arises at the DK phase
transition point \cite{dk,mp}.  In the coupled theory, the
simple saddle point solution described by the Wigner semicircle law
gives a good description of the theory for areas less than the
critical area $A = \pi^2$.  However, above the critical area the
Wigner semicircle violates the constraint $u \leq 1$, and does not
correspond to an acceptable solution.  In the language of
\cite{dk,mp}, this theory is described as a system of  fermions
moving in 1 dimension in an external quadratic potential and
interacting with a logarithmic potential.  When the critical area is
reached, the fermion density is saturated at $h = 0$ and
condensation occurs giving a function $u (h)$ which is identically 1
in a small region.

In terms of the fermion language, we are here considering a system
where the potential is again quadratic but where there is an
impenetrable barrier at the point $h = 1/2$, at which fermions tend to
congregate.  As the area is increased, the strength of the external
quadratic potential increases and the particles move towards the
origin $h = 0$, creating a ``bump'' in $u (h)$.  Eventually, the first
critical point $A_-$ is reached, and the fermion  density reaches
unity at a point $h = e$.   Just as the full theory developed a
condensate where $u = 1$ beyond the critical point, we might expect a
similar effect in the chiral theory above $A = A_-$.  However, unlike
the situation which Douglas and Kazakov considered, our system is not
symmetric, so we cannot expect any particular relation between the
endpoints of the condensation region.
%Those authors
%intuited that the natural way to proceed beyond the single-cut Ansatz
%was to
%allow a set of the eigenvalues to 'condense' along the $u(h)=1$
%border, analogous to the the region $[c,1/2]$ in the solutions
%described in Section 3. Following in this vein,
Thus,
we will adopt an Ansatz for the solution of
(\ref{eq:integral1}) in which  $u (h)$ has support on the region
$(d, 1/2)$, with
\begin{equation}
u(h) = 1 \;\;\;\;\;\; {\rm for}\;\;\;c<h<b\;\;\;{\rm or}\;\;\;a<h<1/2
\label{eq:cond2}
\end{equation}
for some values $d<c<b<a$. Note that $u(h)$ is subject to the
constraint $u(h)<1$ throughout the regions $(d,c)$ and $(b,a)$.

A solution, $u(h)$, satisfying
the double-cut Ansatz (\ref{eq:cond2}) may
be found by adapting the method described in Section 2. The integral
equation for the saddle point thus reads,
\begin{equation}
\frac{Ah}{2}+{\rm ln}{\frac{(h-1/2)(h-b)}{(h-a)(h-c)}} = P \int_R {\rm d}s
\frac{u(s)}{h-s}
\label{eq:integralequation2}
\end{equation}
where the region of integration is $R = (d,c)\cup (b,a)$.
As before, it is useful to define the analytic function
\begin{equation}
 f (h) = \int_R {\rm d}s \frac{u(s)}{h-s}
\label{eq:fdefn2}
\end{equation}
and, letting  $g (h)=\sqrt{(h-a)(h-b)(h-c)(h-d)}$ as appropriate for
the double-cut Ansatz (\ref{eq:cond2}), we find that we may rewrite
$f (h)$ as
\begin{equation}
 f (h) = \frac{Ah}{2}+{\rm ln}{\frac{(h-1/2)(h-b)}{(h-a)(h-c)}} +
g(h)\int_{a}^{1/2}\frac{{\rm d}s}{(h-s) g(s)} +
g(h)\int_{c}^{b} \frac{{\rm d}s}{(h-s) g(s)}
\label{eq:fagain}
\end{equation}
where we have deformed the contours giving terms corresponding to the
pole at $s=h$ and to the logarithmic discontinuities on the intervals
$(a,1/2)$ and $(c,b)$. We may express $f (h)$ from (\ref{eq:fagain})
in terms of elliptic functions. This is particularly useful for
writing $u(h)$ in the region $R$. We find
\begin{eqnarray}
u(h)& = &\frac{2}{\pi \rho} \left[ \sqrt{\frac{(h-c)(h-d)}{(a-h)(h-b)}}
\left((a-b)\Pi(\nu,\frac{(a-d)(h-b)}{(b-d)(h-a)},q)+(h-a)F(\nu,q)\right)
\right.
\nonumber\\
& &\qquad  + \left.
\sqrt{\frac{(a-h)(h-b)}{(h-c)(h-d)}}
\left((c-d)\Pi(\frac{(b-c)(h-d)}{(b-d)(h-c)},q)+(h-c)K(q)\right)\right]
\label{eq:u2}
\end{eqnarray}
where
\begin{equation}
\nu = \sin^{-1}(\sqrt{\frac{(b-d)(1/2-a)}{(a-d)(1/2-b)}})
\;\;\;\;\;\;q=\sqrt{\frac{(b-c)(a-d)}{(a-c)(b-d)}}\;\;\;\;\;
\rho = \sqrt{(a-c)(b-d)}
%\label{eq:}
\end{equation}
and where the notation for the elliptic functions  is that used in
Byrd and Freidman \cite{bf}. Thus $F$,$E$ and $\Pi$ are the
elliptic integrals of the
first, second and third kind, respectively, and when the
angle $\nu$ does not appear as one of the arguments of an elliptic
function
the symbol then represents the associated complete elliptic function
(note $K(q)$ is the complete elliptic function associated to
$F$).

As described in Section 3, the parameters $a,b,c,d$ are not arbitrary
but are subject to conditions related to the definition of $f (h)$ in
(\ref{eq:fdefn2}). As was done in Section 3, we study the asymptotics
of $f$ to determine these conditions on $a,b,c,d$.  Expanding $f(h)$
about $h \rightarrow \infty$ and  isolating the terms of order
${\cal O}(h)$, ${\cal O}(1)$ and ${\cal O}(h^{-1})$ respectively we
find,
\begin{equation}
A = \frac{4}{\rho} ( K(q)-F(\nu,q))
\label{eq:cond2a}
\end{equation}
\begin{equation}
(a-b)\Pi(\nu,\frac{a-d}{b-d},q) + (b-W/2)F(\nu,q) =
(c-d)\Pi(\frac{b-c}{b-d},q)
+ (d-W/2)K(q)
\label{eq:cond2b}
\end{equation}
\begin{equation}
1 = \sqrt{\frac{(1/2-a)(1/2-c)(1/2-d)}{1/2-b}}+\rho(E(q)-E(\nu,q))
+ \frac{(a-b-c+d)^2}{4\rho}(K(q)-F(\nu,q))
\label{eq:cond2c}
\end{equation}
where $W = a+b+c+d$. Note that since there are four unknowns
{$a,b,c,d$}, for fixed area $A$ there will be at most a one parameter
family of solutions to these three equations. Also note that as one
takes $a\rightarrow b$ ({\sl or} $c\rightarrow b$) two of the three
equations above become the constraint equations
(\ref{eq:constraint1a}), (\ref{eq:constraint1b}) of the single-cut
solution.
%In each limit the third equation becomes a point where the single-cut
%solution $u(h)$ is one.

\begin{figure}
\epsfbox{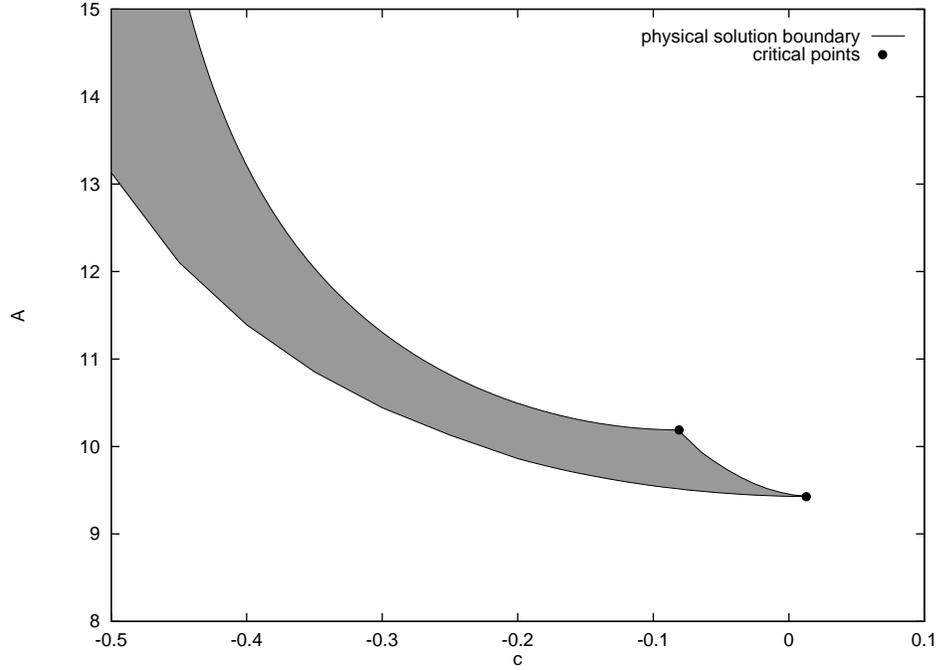}
\caption[x]{\footnotesize double-cut solutions}
\label{f:4}
\end{figure}

The solutions to this double-cut problem which satisfy the constraint
$u (h)\leq 1$ for all $h$ are depicted in Figure~\ref{f:4}.  We find
that there is a continuum of double-cut solutions (the shaded region)
for each $A>A_{-}$.  The upper boundary of the region of solutions is
given by the curve describing the single-cut solution above $A_+$.
The cusp point at the bottom of the region is precisely the lower
critical point. Note that there the value of $c$ is the value of $h$ at
which the ``bump'' in $u$ (see Figure~\ref{f:u})
touches $u (c)= 1$, and not the value
defined in the single-cut solution, which is now taken by $a$ at that
point.

The constraint $u(h)\leq 1$ must be satisfied for all $h$ in order for
the solution $u (h)$ to correspond to a Young diagram at large $N$.
This condition can be used to describe analytically the remaining
boundary curves of the physically acceptable region in
Figure~\ref{f:4}.  As one varies $a,b,c,d$ in such a way as to satisfy
(\ref{eq:cond2a}), (\ref{eq:cond2b}),  and
(\ref{eq:cond2c}),
the solution (\ref{eq:u2}) first fails this condition at the points
$b$ and $c$. Thus, we can find the boundaries of the double-cut
problem simply by studying the slope of the solution $u(h)$ at the
points $b$ and $c$. The boundary of the double-cut solutions arising
from the condition that $u(b+)\leq 1$ is
\begin{equation}
(a-b)(K(q)-F(\nu,q)) = (a-c)(E(q)-E(\nu,q)).
\label{eq:bboundary}
\end{equation}
This is the boundary curve at the bottom of the physical region.
An analysis of the slope of the solution $u(h)$
near the point $c$, ({\it i.e.}, requiring $u(c-)\leq 1$)
yields an equation for the final boundary of the double-cut region.
It reads
\begin{equation}
\rho(b-c)\sqrt{\frac{(1/2-d)(1/2-a)}{(1/2-c)(1/2-b)}} =
(a-c)(c-d)(K(q)-F(\nu,q))-\rho^2(E(q)-E(\nu,q)).
\label{eq:cboundary}
\end{equation}
One finds that the solutions at these boundaries are
indeed legal solutions, so the double-cut region is closed.
It is easy to verify that the $u(c-)\leq 1$
boundary runs precisely between the $A_{+}$ and $A_{-}$.

Having found these new saddle points of the action (\ref{eq:action})
we now study their contribution to the partition function.  To begin
with, we can see from Figure~\ref{f:4} that for all $A > A_-$ there is
indeed a one parameter family of solutions to the saddle point
equation.  However, not all these solutions are actually saddle points,
since the saddle point equation is only satisfied in the interior of
the regions $(d,c)$ and $(b,a)$.  To determine which of the values is
the physical saddle point for any given $A$ it would be necessary to
compute the exact free energy $F$  and to maximize this quantity
across all solutions corresponding to that fixed value of $A$.

Unfortunately, however, an exact calculation of the free energy is
technically difficult.  The situation here is analogous to
that which arises in the full QCD${}_2$ theory.  In that case, it was
shown by Minahan and Polychronakos \cite{mp} that for every value of
$A$ above the critical value $\pi^2$ there is a  family
of solutions to the saddle point equation, parameterized by the
parameter $Q$ corresponding to the $U(1)$ charge, which is essentially
the first moment of $u (h)$.  In that case also,
a direct calculation of the free energy is difficult.  Symmetry
arguments would indicate that the DK solutions with $U(1)$ charge $Q =
0$ are the physical saddle points in this theory.  The correspondence
of the strong coupling expansion of the DK solutions with the
QCD${}_2$ string gives further weight to this conclusion; however
this result has not been rigorously
proven.

In the  chiral theory, we no longer have a reflection symmetry to  rely on;
however, the correspondence of the strong coupling expansion of the
single-cut solution with the chiral string theory is a strong
indication that the free energy is extremized by the single-cut
solutions above $A = A_+$.   We have numerically integrated the free
energy in that region and found that indeed  $F$ seems to be maximized
on the single-cut solution for any fixed $A > A_+$.  The behavior of
the  theory in the intermediate phase, however, presents a much more
subtle problem.  Numerically integrating $F$ indicates that the
physical solutions in the intermediate phase lie near or on
the boundary (\ref{eq:cboundary}).
The  numerical analysis of $F$ in this region is
extremely delicate, presumably due to the existence of phase
transitions at $A_-,A_+$.

Thus, although we have fairly solid evidence that the single-cut
solution describes the physical saddle point above  $A_+$, we are
unable to give an  exact formula for the curve corresponding to the
physical solutions connecting the two critical points.  Nonetheless,
we can compute the derivative of the free energy along the physical
curve using the same method as was used in Section 3 to calculate $F'
(A)$ in the single-cut region.   From the fact that the physical
saddle point obeys the classical equations of motion, we find that the
derivative of the free energy is related to the second moment of the
density function $u (h)$ by
\begin{equation}
F' (A) = -\frac{1}{2}  \int {\rm d} h \; u (h) h^2 + 1/24.
%\label{eq:}
\end{equation}
After some algebra one  ascertains that
\begin{eqnarray}
\lefteqn{\int {\rm d}h \; u(h) h^2}\nonumber \\
 & =& \big(\frac{(1/2-b)(1+W)}{6} + M\big)
\sqrt{\frac{(1/2-a)(1/2-c)(1/2-d)}{1/2-b}}+ M\rho(E(q)-E(\nu,q)) \nonumber
\\
& &+ (3W^2X^2+6WXYZ+Y^2Z^2+2X^2Y^2+2X^2Z^2+X^4)
\frac{(K(q)-F(\nu,q))}{192\rho}
\label{eq:uh2}
\end{eqnarray}
where
\begin{equation}
 W = a+b+c+d\;\;\;\;\;X=a-b-c+d\;\;\;\;\;Y=a+b-c-d\;\;\;\;\;\;Z=a-b+c-d
%\label{eq:}
\end{equation}
and where
\begin{equation}
M = (3W^2+X^2+Y^2+Z^2)/48
%\label{eq:}
\end{equation}
%The expression above is the general expression for the second moment
%of the $u(h)$; we have not used the constraint equations
Using the constraint equations
(\ref{eq:cond2a},
\ref{eq:cond2b}, \ref{eq:cond2c}) allows us to write this expression in
terms of algebraic functions of the area $A$ and the points $a,b,c,d$,
\begin{eqnarray}
\int {\rm d}h u(h) h^2 &=& \frac{(1/2-b)(1+W)}{6}
\sqrt{\frac{(1/2-a)(1/2-c)(1/2-d)}{1/2-b}} + M \nonumber\\
& & +(6WXZY+Y^2Z^2+X^2Y^2+X^2Z^2)A/768
\label{eq:uh22}
\end{eqnarray}
As a simple check note that in the limit $b\rightarrow a$
this expression leads to the result of (\ref{eq:free}).
We have thus given an explicit formula for the derivative of the free
energy in the double-cut region which is applicable to whatever curve
describes the physical solutions in that region, although we have been
unable to describe the curve analytically.
If one could derive an analytic formula for the physical solutions in
the intermediate region,  (\ref{eq:uh22}) could be used to calculate
the order of the phase transition at the two critical points.

\section {Conclusions}
\setcounter{equation}{0}
\baselineskip 18.5pt

% Conclusions/Speculations/Future Directions and Acknowlegments
%

We have developed a matrix model that describes a single
chiral sector of the QCD$_2$ string.  The analysis
indicates that there are three distinct regimes for solutions of the
saddle point equation of the chiral theory on the sphere.  We have
shown that there exists a simple single-cut solution which exists for
large and small areas and which agrees with the string picture in an
expansion around large area.  However, this single-cut solution ends
(stops leading to physically viable solutions) as one decreases the
area from infinity and also as one increases the area from zero.  Thus
the single-cut solution cannot describe the system for areas
$A_{-}<A<A_{+}$.  We have shown that for these intermediate areas a
saddle point may be found as a double-cut solution to the (integral)
equation of motion.  This double-cut solution limits to the single-cut
solution in the large area regime and connects the upper and lower
branches of the single-cut solution.  For any fixed $A>A_{-}$ there is
a one-parameter continuum of saddle point solutions.  When $A > A_+$,
we have fairly conclusive evidence that the single-cut solution
dominates the theory.  However, in the intermediate regime we cannot
calculate exactly where the curve describing the physical saddle point
lies.  In fact, although we believe there is only a single curve with
maximum free energy connecting the two points $A_+$ and $A_-$, it is
conceivable that there may be multiple physical saddle points in this
intermediate region, although numerical evidence indicates that there
is indeed only a single such curve, which lies near or on the
$u(c-)\leq 1$ boundary
(\ref{eq:cboundary}).

Despite our uncertainty regarding the exact location of the
intermediate saddle point curve, our analysis gives a
fairly complete picture of the theory at any $A$. The dominant
stationary solutions are the single-cut solutions
for $A\ge A_{+}$ and $A\leq A_{-}$ and there are likely to
be a simple trajectory of solutions in the double-cut region
that interpolate between these single-cut solutions.
The fact that the physical curve of solutions in $(c,A)$ space, as well
as the solutions $u (h)$, undergo discontinuities at the points $A_+$
and $A_-$ indicates that there are probably phase transitions at these
points.  The behavior of the density function $u (h)$ at the point
$A_-$ is closely analogous to that of the full QCD${}_2$ density
function at the DK phase transition point, suggesting that the chiral
theory probably has a third order phase transition at this point.
However, an exact characterization of  phase transitions in the
chiral theory will not be possible until a method is found for
determining which of the solutions in the double-cut region actually
dominate the partition sum.
If this could be accomplished,
it would then be straightforward in principle to
compute the order of the phase transition at the
critical points simply by comparing power series
expansions of the free energy $F(A)$ on either side
of the critical points.

The single-cut solution of the matrix model gave us an exact analytic
formula for the large area regime of the chiral theory.  Because this
exact formula has a series expansion which is equal to the leading
($N^2$) term in the free energy of the chiral string theory, we were
able to relate the terms in this expansion to sums over maps between
Riemann surfaces.  In particular, a calculation of the leading terms
in the area polynomials of the expansion allowed us to derive a
previously conjectured result on the number of sphere to sphere maps.
An interesting direction for further work might be to find further
structure in the analytic equations for the large area chiral theory
and to relate this structure to the geometry of string maps.

It is remarkable that the location of the critical point
bounding the large area phase is relatively insensitive to the
details of the theory in question.  The result proven here for the
number of sphere to sphere maps was used  in \cite{taylor} to show
that in a chiral theory with branch points, but without
$\Omega$-points, the critical point occurs near $A = 11.9$.  In this
paper we found that the $\Omega$-points in the chiral theory shift the
critical point to near $A = 10.2$, which is extremely close to the DK
critical point in the coupled theory at $A = \pi^2 \approx 9.9$.

One particularly surprising feature of the results given here is the
fact that although the chiral theory apparently has three distinct
phases, the large and small area phases are in fact described by the
same analytic expressions, indicating some type of duality between
these two phases.  This situation also bears an striking similarity to
the results of \cite{taylor} where the convergence properties of the
string expansion were studied for both the chiral and coupled
QCD${}_2$ theories.  In that analysis, it was shown that the string
expansion of the chiral theory has a radius of convergence
approximately equal to the point $A_+$, but that the string series
also converges for very small areas.  A numerical comparison, however,
shows that the results of the matrix model disagree with the explicit
string sum for small area.  The small area phase we have described in
this work of course lies outside the radius of convergence of the
large area phase. Thus, the fact that the string sum did converge for
small $A$ seems mysterious.

In order to further understand the chiral QCD${}_2$ matrix model we
have described here, it would be interesting to study directly the
free energy of this model.  A way of analytically describing the
dominant saddle points for $A_{-}<A<A_{+}$ in this model would be
invaluable for the computation of the order of the phase transition
across the critical loci.  Furthermore, an understanding of how to
compute $F$ directly might shed some light on the significance of the
one-parameter families of solutions described in \cite{mp} for the
full QCD${}_2$ large $N$ theory.

\vskip .5truein
{\Large{\bf Acknowledgements}}

We thankfully acknowledge helpful conversations with
S.\ Axelrod, M.\ Douglas, D.\ Gross, J.\ Minahan, A.\ Polychronakos,
H.\ J. Schnitzer and
I.\ M. Singer. W.\ T.\ would also like to thank the Aspen Center for Physics
where part of this work was completed.

%\begin{equation}

%\label{eq:}
%\end{equation}


\baselineskip 14pt
\begin{thebibliography}{999}
\parindent=.6em
\normalsize
\bibitem{migdal}  A.\  Migdal, {\em Zh.\  Eksp.\  Teor.\  Fiz.\ } {\bf
69}, 810 (1975) (Sov.\  Phys.\  JETP.\  {\bf 42} 413).
\bibitem{rusakov} B.\ Rusakov, {\it
Mod.\ Phys.\ Lett.\ } {\bf A5}, 693 (1990).
\bibitem{witten}  E.  Witten, {\em Comm.  Math.  Phys.} {\bf
141},153 (1991).
\bibitem{gross}  D.\ Gross,   {\it Nucl.\ Phys.\ } {\bf B400},
161 (1993).
\bibitem{minahan}  J.\ Minahan, {\it Phys.\  Rev.\ } {\bf D47},
3430 (1993).
\bibitem{gt1}  D.\ Gross, W.\ Taylor, {\it Nucl.\ Phys.\ } {\bf B400},
181 (1993).
\bibitem{gt2}  D.\ Gross, W.\ Taylor, {\it Nucl.\ Phys.\ } {\bf B403},
395 (1993).
\bibitem{nrs}  S.\ G.\ Naculich,  H.\ A.\ Riggs, H.\ J.\ Schnitzer
{\it Mod.\ Phys.\ Lett.\ } {\bf A8}, 2223 (1993).
\bibitem{dk}  M.\ Douglas, V.\ Kazakov, {\it Phys. Lett.\ } {\bf B319}, 219
(1993).
%Large $N$ phase transition
%in continuum QCD${}_2$, preprint LPTENS-93/20, hep-th/9305047,
%RU-93/17, May 1993.
\bibitem{gm}  D.\ J.\ Gross, A.\ Matytsin,
Instanton Induced Large  $N$ Phase Transitions in Two- and Four-
Dimensional QCD, preprint PUPT-1459, hep-th/9404004, April 1994.
\bibitem{taylor}  W.\ Taylor,
Counting Strings and Phase Transitions in 2D QCD,
preprint MIT-CTP-2297, hep-th/9404175, April 1994.
\bibitem{cmr}  S.\ Cordes, G.\ Moore, and S. Ramgoolam, Large $N$ 2D
Yang-Mills Theory and Topological String Theory,
preprint YCTP-P23-93, hep-th/9402107, February 1994.
\bibitem{horava} P.\ Horava, Topological Strings and QCD in Two
Dimensions, preprint EFI-93-66, hep-th/9311156, November 1993.
\bibitem{rusakov2} B.\ Rusakov, {\em
Phys.\ Lett.\ } {\bf B303},  95 (1993).
\bibitem{mp} J.\ A.\ Minahan, A.\ P.\ Polychronakos,
%Classical Solutions for Two-Dimensional QCD on the Sphere,
%preprint CERN-TH- 7016/93, UVA-HET-93-08, hep-th/9309119,  September 1993.
{\it Nucl. Phys.} {\bf B422} (1994) 172.
\bibitem{cayley} A.\ Cayley, {\em Quart.\ Jnl.\ Pure Appl.\ Math} {\bf
23},  376-378 (1889).
\bibitem{pipkin} A.\ C.\ Pipkin, ``A Course on Integral Equations," Berlin,
Springer, 1991.
\bibitem{bf} P.\ F. \ Byrd, M.D.\ Friedman, ``Handbook of Elliptic
Integrals
for Engineers and Physicists," Berlin, Springer, 1954.

\end{thebibliography}
\end{document}